\documentclass[lineno]{jfm}
\usepackage{graphicx}
\usepackage{newtxtext}
\usepackage{newtxmath}
\usepackage{natbib}
\usepackage{hyperref}
\hypersetup{
    colorlinks = true,
    urlcolor   = blue,
    citecolor  = black,
}

\newcommand{\RomanNumeralCaps}[1]
\linenumbers

\usepackage{bm}
\usepackage{comment}
\usepackage[normalem]{ulem}
\usepackage[amssymb]{SIunits}
\usepackage[dvipsnames]{xcolor}

\title{Collective effects of neighbouring melting ice objects}

\author{Sof\'ia Angriman\aff{1}$\dagger$,
  Detlef Lohse\aff{1,2},
  Roberto Verzicco\aff{3,1},
 \and Sander G. Huisman\aff{1}\footnote[1]{Email addresses for correspondence: s.angriman@utwente.nl, s.g.huisman@utwente.nl}
 \hfill}

\affiliation{
\aff{1}Physics of Fluids Department, Max Planck Center for Complex Fluid Dynamics, and J. M. Burgers Centre for Fluid Dynamics, University of Twente, P.O. Box 217, 7500AE Enschede, The Netherlands
\aff{2} Max Planck Institute for Dynamics and Self-Organization, Am Fa{\ss}berg 17, 37077 G\"ottingen, Germany
\aff{3} Dipartimento di Ingegneria Industriale, University of Rome `Tor Vergata', Roma 00133, Italy \& Gran Sasso Science Institute - Viale F. Crispi, 7 67100 L'Aquila, Italy
}

\begin{document}
\maketitle

\begin{abstract}
We present a study on the melting dynamics of neighbouring ice bodies by means of idealised simulations, focusing on collective effects, with the goal of obtaining fundamental insight into how collective interactions influence the melting of ice.
Two neighbouring (vertically or horizontally aligned), square-shaped, and equally sized ice objects (size on the order of centimetres) are immersed in quiescent fresh water at a temperature of \unit{20}{\celsius}.
By performing two-dimensional direct numerical simulations, and using the phase-field method to model the phase change, the collective melting of these objects is studied. When the objects are horizontally aligned, no significant influence of the neighbouring object on the melting time is observed. On the other hand, when vertically aligned, though the melting of the upper object is mostly unaffected, the melting time and the morphology of the lower ice body strongly depends on the initial inter-object distance. We report that the melting of the bottom object can be enhanced by more than 10\%, or delayed more than 20\%, displaying a non-monotonic dependence on the initial object size. We show that this behaviour results from a non-trivial competition between layering of cold fluid, which lowers the heat transfer, and convective flows, which favour mixing and heat transfer. For this melting in mixed convection, we were able to collapse our data onto a single curve.
\end{abstract}



\section{Introduction}
\label{sec:intro}
Melting is an important and ubiquitous phenomenon present in everyday life, not only when cooling drinks by adding ice cubes, but also engineering applications use phase-change materials for latent-heat storage systems that
optimise energy supply processes \citep{Pielichowska2014, Yang2024_PRXEnergy}. The decrease of the overall global ice mass is a paramount example of melting in the environment. Indeed, accurately quantifying the melting rate of ice bodies like icebergs and packed ice floes is necessary for predicting the impact of melting on ocean dynamics \citep{Cenedese2023}. While icebergs provide a strong motivation, this work focuses on an idealised fresh-water system to isolate the fundamental physical aspects of collective melting.

In natural settings, various mechanisms influence ice melting. For icebergs, volume loss below the surface by buoyant and forced convection can be several times larger than loss due to solar radiation and wind erosion \citep{Savage2001}.
The ice and the surrounding fluid (and flow) are two-way
coupled: the flow enhances the melting, and the cold melt water drives the flow \citep{Du2024}.
Fluid motion may be generated by natural convection, where the velocity in the flow is induced as a result of the buoyancy of the melt, i.e. it is the temperature difference with the ambient what drives motion \citep{Weady2022}.
As a result of the non-monotonic dependence of the density of pure water on temperature, the flow structure and the resulting morphology of melting bodies depends strongly on ambient temperature. The convection may carve sharp pinnacles in the ice, and even dimples, colloquially called scallops \citep{Weady2022, Pegler2020}.
In salty environments such as oceanic icebergs, buoyancy is influenced by both temperature and salinity. When salt is present in the ambient water double diffusive convection may take place. The complex interaction between solutal, thermal, and kinetic boundary layers modifies the ice surface, which can also give rise to scallop formation \citep{Wilson2023, Yang2023b, Xu2025}.
The presence of an external flow, resulting in forced convection, has also been reported to modify the melting dynamics \citep{Hao2002, Yang2024_JFM}.
Other factors such as  bubbles and sediments released from ice \citep{Wengrove2023}, the aspect ratio of ice bodies \citep{Hester2021, Yang2024, Yang2024_JFM}, or vertical oscillations in cold, salty environments \citep{Sweetman2025}, have been shown to affect flow, morphology, and overall melting rates. 

Nearby melting ice objects is an additional factor to consider. 
The interaction between multiple sources of subglacial discharge has been shown to modify the melting of glaciers in a non-trivial way  \citep{Cenedese2014, Cenedese2016}, highlighting the importance of considering the interplay of plumes in melting phenomena. The complex structure of ice in natural settings, like frazil ice and ice floes which can melt, collide, or raft on top of each other provides further motivation for evaluating collective effects of melting bodies \citep{Golden2020, Banwell2023, Opfergelt2024}.
In systems where convection is driven by gradients of concentration\textemdash{}instead of by temperature or salinity differences\textemdash{}collective effects have shown to be important, as is the case of droplets dissolving close to each other \citep{Chong2020}.
In these systems not only the ambient conditions are important, but also the interaction between the plumes of the objects, and how this feedbacks on the flow, should be taken into account. In the context of latent heat storage systems which use phase change materials, the design of the devices is key for maximising efficiency \citep{Groulx2021, Yang2024_PRXEnergy}. Understanding fundamentals of the interaction of objects that change phase can then contribute to design optimisation.

In this study, we address collectivity in ice melting in an idealised system through direct numerical simulations. We consider two identical ice objects, fully immersed in fresh water, and explore both vertical and horizontal alignment while varying the separation and object size.
Our goal is to isolate how the inter-object spacing modifies the melting and flow dynamics, with the aim of providing fundamental insight into how collective interactions affect ice melting.
To mimic standard lab conditions, the initial ambient temperature will be set to  $20\celsius$. In this way we mostly avoid the additional complexity of the density inversion, as its effect will be weak.

The paper is organised as follows: In section \ref{sec:setup} we describe the numerical setup, presenting the equations of motion and the range of parameter space explored. In \S\ref{sec:melting_results} we compare the melting times of the top and bottom ice objects. Section \ref{sec:phenomenology} presents a discussion on the morphology of the ice, and the observed phenomenology. Finally, we draw the conclusions in \S\ref{sec:conclusions}, and present perspectives for future research.
 
\section{Numerical setup}
\label{sec:setup}

\begin{figure}
\centering
\includegraphics[width=\textwidth]{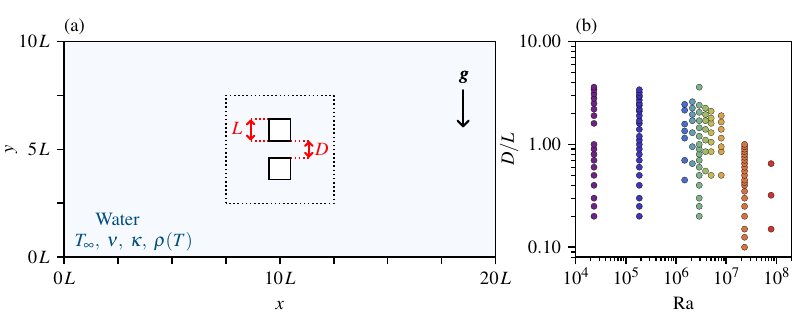}
\hfill
  \caption{(a) Simulation setup. Two ice objects of size $L$ are placed one on top of the other at a distance $D$. The dashed lines indicates the zoomed-in region for visualisation. (b) Region of parameter space explored in this work. The colour code corresponds to $\text{Ra}$.}
\label{fig:setup}
\end{figure}

We perform two dimensional (2D) direct numerical simulations (DNS) of two melting ice objects.
For Prandtl numbers larger than 1, heat transfer in 2D and 3D Rayleigh\textendash{}Bénard convection has been shown to be nearly identical  \citep{vanderPoel2013}. 
Furthermore, qualitative and quantitative agreement between 2D and 3D numerical simulations of ice melting has also been reported \citep{Purseed2020, Hester2021, Yang2023, Yang2024_JFM}, so we also restrict ourselves to 2D simulations, allowing us to explore higher Rayleigh numbers.
We consider initially square objects of size $L$, at a temperature set equal to their melting temperature $T_m = \unit{0}{\celsius}$. They are immersed in a closed, adiabatic container of size $20L$ in the horizontal direction, and $10L$ in the vertical direction (this being the direction of gravity ${\bm g}$) filled with fresh, quiescent water initially at temperature $T_\infty = \unit{20}{\celsius}$ (note that by the end of the evolution the water temperature is about $\unit{19}{\celsius}$). The objects are stacked vertically at a distance $D$, symmetrically with respect to the centre of the domain. The setup is schematically shown in figure~\ref{fig:setup}(a).

To account for the maximum in density that fresh water displays at $T_\text{max} = \unit{4}{\celsius}$ we consider a non-linear equation of state \citep{Gebhart1977}, relating density $\rho$ and temperature $T$, 
\begin{equation}
    \rho(T) = \rho_0(1 - \beta |T - T_\text{max}|^q),
\end{equation}
where $\rho_0$ is a reference density and $\beta = \unit{9.3\times 10^{-6}}{\kelvin^{-q}}$ is the generalised thermal expansion coefficient with exponent $q=1.895$. The effect of the density anomaly will be weak, as the ambient temperature remains sufficiently larger than  $T_\text{max}$ throughout the whole evolution.
To model the phase change we consider the phase-field method (based on \citet{Yang2024, Favier2019, Liu2021}), where a scalar field $\phi$ that transitions smoothly from $0$ in the liquid to $1$ in the solid is evolved in time and space. The position of the interface is then defined implicitly where $\phi = 0.5$. 
Hence, the equations to solve correspond to the incompressible Navier\textendash{}Stokes equations within the generalised Boussinesq approximation, and advection-diffusion equation for the temperature, and an equation for $\phi$. These read, in non-dimensional form,
\begin{equation}
    \centering
    \begin{cases}
        \frac{\partial}{\partial t} \bm{u} + (\bm{u} \cdot \bm{\nabla}) \bm{u} &= -\bm{\nabla} p + \sqrt{\frac{\text{Pr}}{\text{Ra}}}\Big (\nabla^2 \bm{u} - \frac{\phi \bm{u}}{\eta}\Big) + |\theta - \theta_\text{max}|^q\, \hat{y}\\
        \frac{\partial}{\partial t} \theta + (\bm{u} \cdot \bm{\nabla}) \theta &=  \frac{1}{\sqrt{\text{Ra Pr}}}\nabla^2 \theta + \frac{1}{\text{Ste}}~ \frac{\partial}{\partial t} \phi\\
        \frac{\partial}{\partial t} \phi &= \frac{6}{5} \, \frac{\text{Ste}}{ C\,  \sqrt{\text{Ra Pr}}} \Big[\nabla^2 \phi - \frac{1}{\varepsilon^2} \phi(1 - \phi)(1 - 2 \phi + C\theta)\Big].
    \end{cases}
    \label{eq:NS_eqs}
\end{equation}
Here $\theta = (T-T_m)/(T_\infty-T_m)=(T-T_m)/\Delta T$ is the non-dimensional temperature difference, $\bm{u}$ is the incompressible velocity field (i.e. $\bm{\nabla} \cdot \bm{u} = 0$) in units of the free-fall velocity $U_f = \sqrt{\beta g \Delta T^q L}$, and $p$ is the non-dimensional pressure, using $\rho_0 U_f^2$ as the reference pressure. The relevant non-dimensional numbers of the problem correspond to the Rayleigh number $\text{Ra}$, the Prandtl number $\text{Pr}$, and the Stefan number $\text{Ste}$, here defined as
\begin{equation}
      \text{Ra} = \frac{g \beta \Delta T^q L^3}{\nu \kappa},
\quad \text{Pr} = \frac{\nu}{\kappa},
\quad \text{Ste} = \frac{c_p \Delta T}{\mathcal{L}},
\end{equation}
where $\nu$ and $\kappa$ are the kinematic viscosity and thermal diffusivity, respectively (assumed to be equal in the solid and liquid phases), $c_p$ is the specific heat capacity, and $\mathcal{L}$ is the latent heat of melting. In eqs.\eqref{eq:NS_eqs} $\varepsilon$ is the non-dimensional diffuse interface thickness in units of $L$, and typically set to the mean grid spacing. The variable $\eta$ is the non-dimensional damping time, the typical time over which the fluid velocity is forced to the solid velocity within the object (in our study, the solid is at rest), which is set to $\Delta t$, the time-step in units of $L/U_f$, following \citet{Favier2019}.
$C$ is the phase mobility parameter, which arises from the Gibbs-Thomson relation linking the interface temperature and its curvature, and here fixed at $C=10$ (see more details in \cite{Yang2024, Hester2020}). Equations \eqref{eq:NS_eqs} are solved using the second-order, staggered, finite-difference solver AFiD \citep{Verzicco1996,Ostilla2015,VanderPoel2015}, with no-slip boundary conditions on all sides. Since the domain is adiabatic, for the temperature field we impose Neumann boundary conditions.
The velocity and temperature fields are solved in a uniform mesh in both directions. The number of grid points are such that there are at least 3 grid points in the thermal boundary layer, which scales as $\delta_\text{th} \propto L~\text{Ra}^{-1/4}$ \citep{Bejan1993}. The phase field $\phi$ is solved in a refined mesh, using twice as many divisions in both directions as the mesh where temperature and velocity are solved. We verified that this resolution was sufficient to accurately resolve the melting dynamics. A grid convergence study is discussed in appendix~\ref{sec:grid_convergence}.

We consider objects whose initial size $L$ ranges from $0.5~\text{cm}$ to $7.5~\text{cm}$, and we fix $\text{Pr} =7$ and $\text{Ste} = 0.25$ (corresponding to water at $T_\infty = \unit{20}{\celsius})$, thus spanning Rayleigh numbers in the range $10^4 \leq \text{Ra} \leq 10^7$.
For all cases, the initial ice-to-water area ratio is the same $A_\text{ice}/A_\text{water} \approx 0.01$.
For each size, we set different initial displacements $D/L$. Figure \ref{fig:setup}(b) summarises the parameter space explored in this work. For each object size, we also run an additional case to use as reference, by placing a single object of size $L$ in the centre of the domain (i.e., with its centre located at $x/L = 10$ and at $y/L = 5$).

\section{Results}
\label{sec:results}
\subsection{Melting times of top and bottom objects}
\label{sec:melting_results}

\begin{figure}
\centering
\includegraphics[width=1\textwidth]{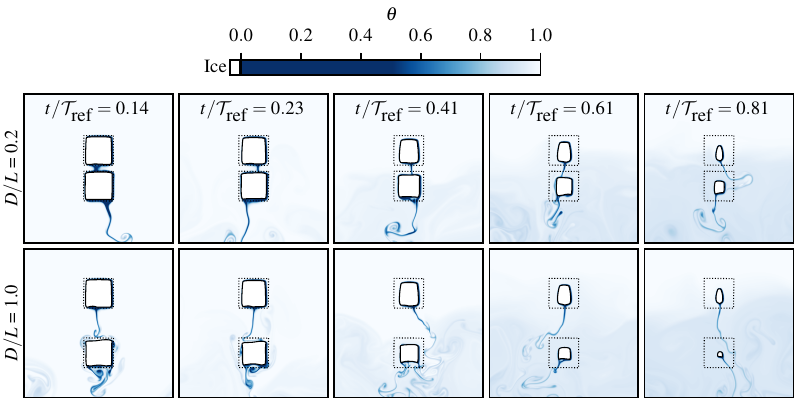}
\hfill
  \caption{Zoomed-in evolution (see figure~\ref{fig:setup}a) of the melting of the top and bottom objects with initial size $L
  = 5~\text{cm}$ ($\text{Ra} \approx 2.3\times 10^7$). The dotted lines indicate the initial contours. The instantaneous snapshots of the non-dimensional temperature field $\theta$ are shown for different times, in units of the reference case melting time $\mathcal{T}_\text{ref}$. 
The colour-bar indicates $\theta$ within the liquid, and the region corresponding to ice is shown in white.
  The top row corresponds to an initial displacement $D/L = 0.2$ ($D/\delta_\text{th} \approx 14$), where the bottom melts slower than the top. The bottom row is for $D/L = 1.0$ ($D/\delta_\text{th} \approx 69$), where the  object below melts faster.}
\label{fig:snapshots}
\end{figure}
We begin with a qualitative analysis of the melting behaviour of the top and bottom objects as the initial distance between them varies. Figure \ref{fig:snapshots} shows snapshots at different times of the evolution of objects 
with an initial Rayleigh number $\text{Ra} \approx 2.3\times 10^7$ ($L = 5~\text{cm}$).

When comparing the overall evolution of the top object in both configurations (small and large $D$), no noticeable difference in its morphology and melting time is observed. Note that the shape of the top object is similar for fixed times (column-wise) for both distances. On the other hand, the lower body displays significant differences between the two configurations. For small $D$ (top row), the ice at the bottom melts slower than the top one. 
The melt water plume from the top object is mostly laminar, and it remains attached to the object below for a large part of its evolution. For large $D$ (bottom row) we observe an unstable, oscillating plume, with a horizontal spread comparable to the object size, increasing mixing and circulation (note the eddies present in the bottom side of the bottom ice), and causing a higher melt rate of the bottom object.

\begin{figure}
\centering
\includegraphics[width=1\textwidth]{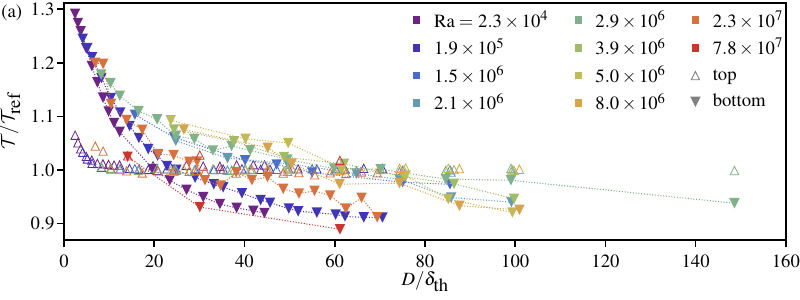}
\hfill
\includegraphics[width=1\textwidth]{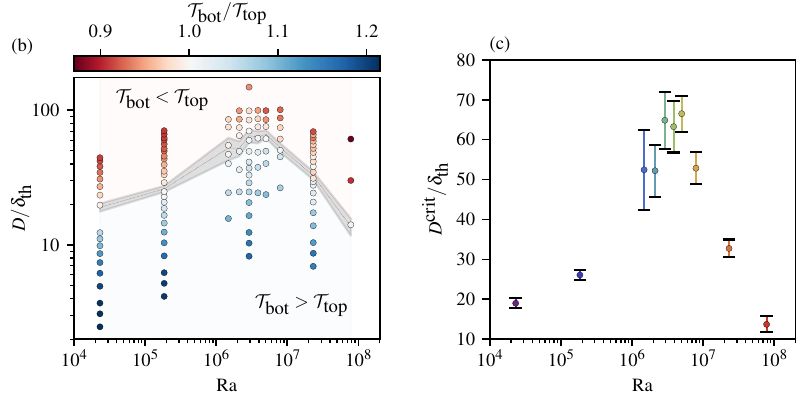}
\hfill
  \caption{(a) Melting times $\mathcal{T}_\text{top}$ and $\mathcal{T}_\text{bot}$ of the top (open, pointing-up markers) and bottom objects (filled, pointing down markers), respectively, in units of the reference melting time $\mathcal{T}_\text{ref}$, as a function of the vertical distance between objects $D$, for different $\text{Ra}$ (i.e. for different initial object sizes), indicated by the different colours. The distance is normalised by the estimation of the thermal boundary layer based on the initial object size $\delta_\text{th} \propto L~\text{Ra}^{-1/4}$.
  (b) 2D map of the ratio of bottom to top melting times.
  The shaded grey region indicates the critical distance $D^\text{crit}$ where the top and bottom melting times are equal. The blue (red) symbols correspond to distances where $\mathcal{T}_\text{bot}$ is greater (smaller) than $\mathcal{T}_\text{top}$.
  (c) Critical distance $D^\text{crit}$ and its uncertainty, normalised by $\delta_\text{th}$, as a function of $\text{Ra}$. Colours as in panel (a).
  }
\label{fig:melting_times}
\end{figure}

This top-bottom asymmetry in the melting times for the objects is systematic and is present for all the object sizes considered. Figure~\ref{fig:melting_times}(a) shows the melting times $\mathcal{T}_\text{top}$ and $\mathcal{T}_\text{bot}$ of the top and bottom object, respectively, normalised by $\mathcal{T}_\text{ref}$, as a function of the initial distance, for all of the object sizes (i.e. initial $\text{Ra}$). Note that in order to compare between the different sizes, the inter-object distance is normalised by the estimation of thermal boundary layer thickness based on the initial object size $\delta_\text{th} \propto L~\text{Ra}^{-1/4}$.
We chose this length scale for normalising the distance as it has a more relevant physical meaning than the object size itself: it is the typical distance over which the temperature varies from the melting temperature to the ambient temperature.
In the appendix~\ref{sec:area_nusselt} we show the temporal evolution of the area of the ice bodies, and of the Nusselt number $\text{Nu}$, the non-dimensional heat flow. A reasonable scaling $\text{Nu} \propto\text{Ra}^{1/4}$ is reported, consistent with what is observed for heat transfer in natural convection \citep{Bejan1993}. Some deviation from this scaling is observed, as the inter-object distance is not accounted for. We will go back to this point later on this section.

For inter-object distances $D/\delta_\text{th} \gtrsim 5$, the ratio of top-to-reference melting times $\mathcal{T}_\text{top}/\mathcal{T}_\text{ref} \approx 1$, implying that the top object's melting time is not affected by the presence of the object below. This similarity with the reference is also observed when comparing the evolution of its morphology.
When the objects are very close together a slight increase in $\mathcal{T}_\text{top}$ is observed, as a result of an accumulation of cold melt water in between the two bodies which decreases the heat transfer.
On the other hand, the melting time of the bottom object displays a dramatic dependence on $D$, for all $\text{Ra}$ explored. For small enough distances ($D/\delta_\text{th} \lesssim 20$), $\mathcal{T}_\text{bot} > \mathcal{T}_\text{top}$, while when the inter-object distance increases the melting of the bottom object is enhanced and thus $\mathcal{T}_\text{bot} < \mathcal{T}_\text{top}$. This behaviour in $\mathcal{T}_\text{bot}$ is robust to perturbations of the initial conditions.
Variations in the melting time can be estimated by performing several realisations of the same configuration, initialising the background temperature field from a continuous uniform distribution $\unit{19}{\celsius}\leq T \leq \unit{21}{\celsius}$ for each grid cell. The maximum variation in the bottom melting time $\mathcal{T}_\text{bot}$ is of order $2\%$, reflecting the chaotic nature of the flow, and the variations in melting times are larger for the cases where $D$ is larger. The typical variations of $\mathcal{T}_\text{top}$ are of $0.5\%$.
We also studied the effect of varying the inter-object distance horizontally. Differences in the melting time with respect to the single object case are of the order of $3\%$, and we did not observe significant variations between the different distances considered. More details are discussed on appendix~\ref{sec:horizontal}.

The ratio of top and bottom melting times for all of the datasets is shown in panel (b) of figure~\ref{fig:melting_times}.
Again, two regimes are identified: for small values of $D$, the melting of the bottom object is delayed and up to 22\% longer, while when the distance increases bottom melting is enhanced by up to 10\%. We observe that $D^\text{crit}$, the distance at which $\mathcal{T}_\text{bot}/\mathcal{T}_\text{top} = 1$, varies with $\text{Ra}$, and it does so in a non-monotonic way. 
In order to quantify this dependence, we consider the (purely empirical) relation between the ratio of melting times and the inter-object distance, 
\begin{equation}
\mathcal{T}_\text{bot}/\mathcal{T}_\text{top} = a~ (D/\delta_\text{th}^\text{ini})^\alpha + b,
\label{eq:fit}
\end{equation}
which yields $D^\text{crit}/\delta_\text{th} = [(1-b)/a]^{1/\alpha}$ for the critical distance.
We then performed several fits of the data using expression~\eqref{eq:fit}, sampling each point from a Gaussian distribution $\mathcal{N}(\mu, \sigma^2)$, where $\mu$ corresponds to the measured data, and $\sigma$ to the typical variation in different realisations, as previously discussed. From each fit we obtain a value of $D^\text{crit}$, and then compute the mean value and a standard deviation from all of the sets of data. The result of this is shown in figure~\ref{fig:melting_times}(c), where a clear non-monotonic dependence on $\text{Ra}$ is observed (and is also shown in panel (b) with the shaded grey area). The critical distance first increases, it reaches its maximum value at around $\text{Ra} \approx 3 \times 10^6$, and then decreases again. Interestingly, for the Rayleigh numbers close to where $D^\text{crit}$ peaks, the error-bar is larger, implying that for these configurations the system displays a higher sensitivity to initial conditions. As it will be discussed by the end of section \ref{sec:phenomenology}, this range of $\text{Ra}$ corresponds to the observed transition from a 
stable to a more unstable, fluctuating plume. 

\begin{figure}
\centering
\includegraphics[width=1\textwidth]{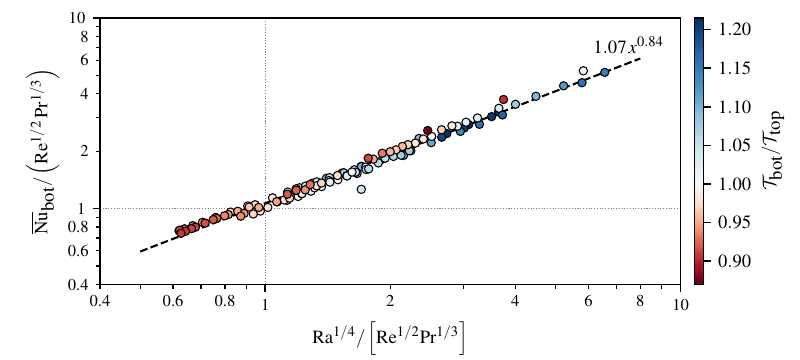}
\hfill
  \caption{
  Average Nusselt number of the bottom object normalised by the forced convection heat transfer scaling. The points are coloured according to the ratio of melting times $\mathcal{T}_\text{bot}/\mathcal{T}_\text{top}$, indicated by the colour bar. The typical error bar is of the size of the marker. The line corresponds to a fit of the data with the function $y = b\, x^a$, yielding $a = 0.84 \pm0.02$, and $b = 1.07\pm0.01$ (95\% confidence intervals).
  }
\label{fig:Nubot_mixed_convection}
\end{figure}

Further insights into the melting time of the bottom object can be gained within the framework of mixed convection, i.e. the combined effect of natural and forced convection. Following \citet{Bejan2013}, for heat transfer in mixed convection, we can define a group that compares the relative strength of natural and forced convection. Under natural convection, heat transfer scales proportional to $\text{Ra}^{1/4}$ \citep{Bejan1993, Yang2024}, while in forced convection it is proportional to $\text{Re}^{1/2} \text{Pr}^{1/3}$ for large $\text{Pr}$ \citep{Bejan1993, Yang2023}. The parameter $\text{Re} = u \ell/\nu$ is the Reynolds number (with $u$ the typical velocity of the external flow that sets the forced convection, and $\ell$ a relevant length-scale).
Then, the group
\begin{equation}
    \frac{\textrm{Ra}^{1/4}}{\textrm{Re}^{1/2}\, \textrm{Pr}^{1/3}}
    \label{eq:mixed_conv}
\end{equation}
determines if forced or natural convection dominates the dynamics.
When the quantity in $\eqref{eq:mixed_conv}$ is smaller than unity, then forced convection dominates, while when it is greater than 1 it is natural convection the mechanism that wins.
For our specific configuration, we use $\ell = D$, and take $u$ as the typical velocity of the plume of the single object at a distance $D$. We then compute the non-dimensional surface-averaged heat flow of the bottom object (see  appendix~\ref{sec:area_nusselt} for a detailed derivation),
\begin{equation}
    \overline{\text{Nu}}_\text{bot} = \frac{L^2/\kappa}{\mathcal{T}_\text{bot}}\, \frac{1}{\text{Ste}}.
\end{equation}
In figure \ref{fig:Nubot_mixed_convection} we plot $\overline{\text{Nu}}_\text{bot}$ compensated by $\text{Re}^{1/2}\,\text{Pr}^{1/3}$, the heat transfer scaling corresponding to forced convection. 
We observe that this normalisation nicely collapses the data across all of the simulations onto a single curve. 
For small values of the control parameter, this ratio should plateau at a fixed value, indicating that heat transfer is due solely to forced convection, while for large values the ratio should have a slope proportional to $1$ (i.e. $\overline{\text{Nu}} \propto \text{Ra}^{1/4}$ irrespective of $\text{Re}$). In our simulations the control parameter from expression ~\ref{eq:mixed_conv} is of order one, which indicates that melting indeed occurs as a result of the combination of forced and natural convection. Furthermore, by fitting the data with power law, an exponent of $0.84 \pm 0.02$ is obtained, a value in between $0$ and $1$ consistent with the limits of purely forced convection and purely natural convection previously discussed.
In the present study, the bottom object melts as a result of its own natural convection and from the flow generated by the melt of the top object, which is time dependent. Interestingly, in spite of this, a description that does not directly account for the time dependence of the forced convection source captures the global behaviour of the melting time of the lower object.

\subsection{Phenomenology}
\label{sec:phenomenology}
\begin{figure}
\centering
\includegraphics[width=1\textwidth]{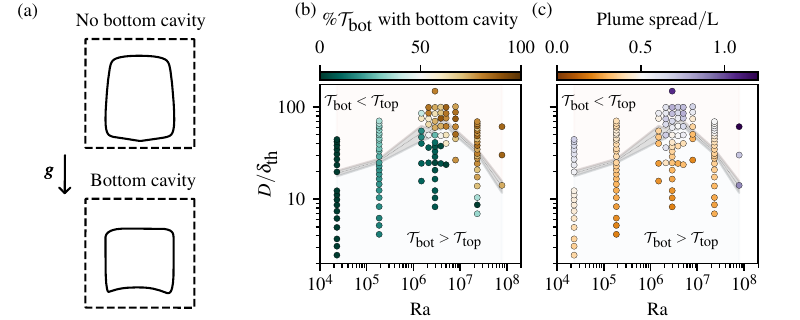}
\hfill
  \caption{
  (a) Representative contours of the ice morphology, for cases without and with a cavity on the lower face of the object (``bottom'' cavity).
  The profiles corresponds to $\text{Ra} \approx 5.0 \times 10^6$ at $t\approx 0.43~\mathcal{T}_\text{ref}$ for $D/\delta_\text{th} \approx 85.2$. The dashed lines indicate the initial ice contour, shown for reference.
  (b) Percentage of the total bottom object's evolution where a cavity on its bottom face is present. 
  (c) Typical horizontal amplitude (in units of the object size $L$) spanned by the plume of the reference case at a given distance $D$.
  In panels (b) and (c) the grey shaded region indicates the distance where $\mathcal{T}_\text{bot} = \mathcal{T}_\text{top}$. The regions where $\mathcal{T}_\text{bot}$ is larger and smaller than $\mathcal{T}_\text{top}$ are also indicated.
  }
\label{fig:cavity_plume}
\end{figure}

We first focus on the regime in which $\mathcal{T}_\text{bot} > \mathcal{T}_\text{top}$. As seen in the top row of figure~\ref{fig:snapshots} (corresponding to a small distance, $D/\delta_\text{th} \approx 14$), the melt water plume in which the bottom object sits in is mostly laminar: by the time it reaches the object below, the plume does not display significant flapping, and the cold melt water flows around the lower block, remaining attached to its surface.
This flow continues along the underside of the bottom object, in close contact to the surface. The resulting layer of cold water acts as thermal shielding, which explains why $\mathcal{T}_\text{bot}/\mathcal{T}_\text{top} > 1$.
Moreover, for the smallest distances considered the cold melt water also accumulates in the region in between the two bodies, further preventing the melting. 

When the objects are farther apart, and $\mathcal{T}_\text{bot}$ is smaller or of the order of $\mathcal{T}_\text{top}$, a cavity forms at the bottom of the object below. This is shown in figure~\ref{fig:snapshots}, where the lower object develops a cavity on its bottom face when $D/L = 1$ (or equivalently $D/\delta_\text{th} \approx 69$); see for instance the snapshot at time $t=0.41~\mathcal{T}_\text{ref}$. A contour of the bottom object when it has developed a cavity below is also shown in panel (a) in figure~\ref{fig:cavity_plume}.
The contour of the upper object, which does not develop a bottom cavity, is shown for comparison.
The presence of a cavity on the underside, that persists for a large part of the evolution of the bottom object, correlates well with enhanced melting. Figure~\ref{fig:cavity_plume}(b) shows a 2D map of the percentage of the evolution of the object below where a bottom cavity is detected, for all $\text{Ra}$ and all $D$ considered. 
To detect the presence of a bottom cavity we compute the convex hull of the lower half part of the ice, and compare its area with the actual surface of the actual portion of ice. When the convex hull area is greater than the ice surface by 1\% a bottom cavity is detected.
For large enough objects (here, $\text{Ra} > \mathcal{O}(10^6)$), a persistent bottom cavity (that is, a cavity that exists for more than half of the object's evolution) correlates well with $\mathcal{T}_\text{bot} < \mathcal{T}_\text{top}$, i.e. with the enhanced melting.
This effect can also be observed in the movies provided as supplementary material.
Neither the top objects nor the single melting object develop a cavity on their underside. 
As noted by \cite{Yang2024}, sufficiently large ice bodies are expected to form a lower cavity due to enhanced circulation from flow separation. For smaller objects (i.e., smaller $\text{Ra}$), the flow produced by natural convection is not strong enough to separate at the corners of the ice, preventing the formation of a cavity. As $\text{Ra}$ increases, the buoyancy-driven flow intensifies and flow separation can occur underneath sufficiently large bodies, triggering cavity formation.
In our case, the mechanism is similar as that reported by \cite{Yang2024} in that separation plays a key role, but differs in that the circulation beneath the lower object results from a combination of flow from its own meltwater  (natural convection) and the forced convection produced by the impinging plume from the top block.

When a bottom cavity forms, the plume shed from the top object is oscillating from side to side, spanning horizontally a region that encompasses the bottom object. The downdraught of the meltwater plume triggers flow separation, which forces a circulation on the underside of the lower object. This carves the ice due to enhanced mixing, and results in the formation of the cavity.
To quantify the plume spread, we consider the single object case, and estimate the typical width of the region spanned by the plume as it oscillates, as a function of the distance to the object. Figure~\ref{fig:cavity_plume}(c) shows the typical width of the plume, in units of the object size $L$, at each of the initial displacements considered. A plume oscillating with a typical amplitude of half the initial object size $L/2$ shows good correlation with the enhanced melting regime $\mathcal{T}_\text{bot} < \mathcal{T}_\text{top}$.
For $\text{Ra} > 1.9 \times 10^5$, the Pearson correlation coefficient between the persistence of a bottom cavity and the plume spread is $\rho \approx 0.51$ (see figure~\ref{fig:correlations}(a) in appendix~\ref{sec:correlations} for a supporting graph).
Even though the plume produced by the top object is slightly modified by the presence of its neighbour below, it is interesting that by analysing the properties of the plume of the single object case we find an observable that is linked to the enhanced melting. This is an important point when considering more complex configurations, or different object geometries.

\begin{figure}
\centering
\includegraphics[width=1\textwidth]{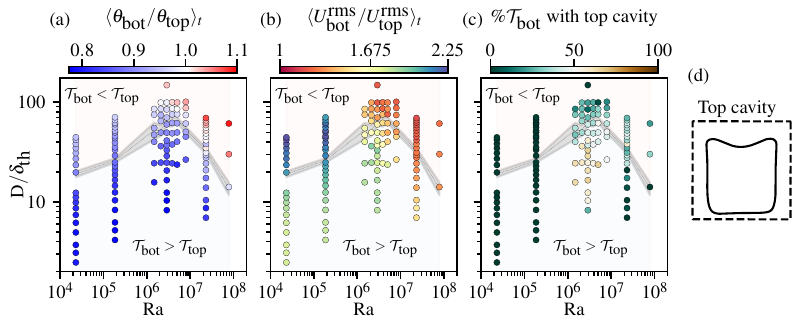}
\hfill
  \caption{
  (a-b) Temporal average of ratio of bottom to top temperature, and root mean square velocity, respectively, around each object.
  (c) Percentage of the bottom object's evolution where a top cavity is present. 
  (d) Representative contour of the morphology of the bottom object when it develops a cavity on its upper side. The contour corresponds to $\text{Ra} \approx 5.0\times 10^6$ when $D/\delta_\text{th}\approx 23.7$ at $t/\mathcal{T}_\text{ref} \approx 0.43$. The dashed line shows the initial ice contour.
  In panels (a-c) the grey shaded region indicates the distance where $\mathcal{T}_\text{bot} = \mathcal{T}_\text{top}$. The regions where $\mathcal{T}_\text{bot}$ is larger and smaller than $\mathcal{T}_\text{top}$ are also indicated.
  }
\label{fig:avg_quantities}
\end{figure}

Note that for the two smallest $\text{Ra}$ considered the object below does not develop a cavity for the distances where bottom melting is enhanced, as a result of the forcing from the top  plume not being strong enough to trigger flow separation, as the objects are too small (small Reynolds numbers).
To gain further understanding of the melting dynamics, we compute the ratio of the average temperature of the fluid, and of the root mean square of the flow velocity in the surrounding of each object, considering points within $3\delta_\text{th}$ from the ice boundary, to serve as a proxy of local gradients at the interface,
\begin{equation}
    \Big\langle \frac{\theta_\text{bot}}{\theta_\text{top}}\Big\rangle_t ,\qquad
    \Big\langle \frac{U^\text{rms}_\text{bot}}{U^\text{rms}_\text{top}}\Big\rangle_t,
\end{equation}
where $\langle \cdot \rangle_t$ indicates an average in time. These ratios are shown in panels (a) and (b) of figure~\ref{fig:avg_quantities}.
By looking at the ratio of temperatures in figure~\ref{fig:avg_quantities}(a),
we observe that for $\text{Ra} > \mathcal{O}(10^6)$, the enhanced melting regime corresponds to the bottom object being surrounded by water which is, on average, warmer than that at around the top.
The origin of the temperature increase is linked to the spread of the top plume. Indeed, $\langle\theta_\text{bot}/\theta_\text{top}\rangle_t$ is correlated with the plume spread with a Pearson correlation coefficient $\rho \approx 0.69$; these data are shown in figure~\ref{fig:correlations}(b) in appendix~\ref{sec:correlations}. This indicates that warm water entrainment in the vicinity of the bottom object is related to the breadth of the top plume in which the object is immersed.
For the objects with the smallest Rayleigh number, with $\text{Ra} \lesssim 10^6$, the surroundings of the lower body are always colder than at the top.
When analysing the ratio of velocities in panel (b) of figure~\ref{fig:avg_quantities} we first note that this ratio is always greater than one, indicating that the object below is immersed in a flow with higher fluctuations for all configurations explored. 
Focusing on the two smallest $\text{Ra}$, as the inter-object distance increases, fluctuations are up to $25\%$ larger than at the smallest value of $D$ explored.
Then, we link the accelerated melting of the object to this increase in velocity fluctuations. The enhanced melting occurs even though the surroundings are at a lower temperature and no flow separation is observed for these object sizes.
For these smaller $\text{Ra}$, the flow established in the box remains relatively localised around the horizontal position of the ice blocks $x=10L$ (see movies), which leads to a stronger flow around the lower object at larger separations.
For $\text{Ra} > 10^6$, the data show that the fluctuations around the bottom object grow as the distance to its top neighbour decreases, even though melting is delayed as $D$ decreases. This is presumably linked to the flow becoming less localised, with a large-scale circulation developing that affects both ice bodies more uniformly, reducing the velocity contrast between them. This highlights that the melting time and the morphology of the ice body below depends on a non-trivial competition between the shielding effect of cold melt water, and convective flows which favour mixing and heat transfer.

For the intermediate $\text{Ra}$ considered, for small $D$ a cavity forms at the top of the bottom object, where the cold water plume impacts. 
A representative contour of the bottom ice block when it develops a cavity on its upper side is shown
in figure~\ref{fig:avg_quantities}(d), for an object with $\text{Ra} = 5.0 \times 10^6$ (i.e. $L = 3$~cm), and where the initial distance to the top object is $D/\delta_\text{th} \approx 24$ (see also the movies provided as supplementary material). 
As before, we estimate how persistent in time this top cavity is, shown in figure~\ref{fig:avg_quantities}(c). For the range of $\text{Ra}$ around the peak of $D^\text{crit}$ the formation of a persistent top cavity correlates well with $\mathcal{T}_\text{bot} > \mathcal{T}_\text{top}$.
The top cavity reflects the higher $D^\text{crit}$ for these object sizes.
This Rayleigh number range corresponds to the destabilisation of meltwater plume of the top body, where stronger bulk flow fluctuations are observed. The plume, while not oscillating significantly, carries enough momentum to locally enhance melting and carve the cavity on the upper side of the lower ice block. Despite this local enhancement of melting, cold water remains attached to the surface and accumulates below, so overall, the shielding effect dominates, slowing down the melting.

\section{Conclusion}
\label{sec:conclusions}
We have studied the collective effects of the melting of two water-immersed and vertically aligned ice bodies. By performing numerical simulations, varying the initial displacement and considering different  Rayleigh numbers (corresponding to changing the objects' size), we have shown that the melting time of the lower object drastically varies, depending on the initial distance to the body above. When the objects are close together, the cold water plume that flows from the upper object onto the lower one is laminar, which shields the bottom object and extends its melting time by more than $20\%$. 
When the distance increases,
the plume is unstable
and we observe flow separation, leading to a circulation that increases the local heat flux, accelerating the melting by more than $10\%$.
We linked this enhancement to the formation of a cavity on the lower face of the bottom object. We observed that in the configuration explored in this work, a good criterion for predicting whether melting will be enhanced or delayed is to quantify the typical horizontal spread of the plume of a single object (the reference case considered here). This provides hints for the expected behaviour when considering different object shapes, and possibly other configurations.

We observed that the two reported regimes of enhanced and delayed melting of the bottom object are present for all the object sizes considered. However, the inter-object distance, when measured in units of $\delta_\text{th}$, at which the transition between these two behaviours occurs has a strong, non-monotonic dependence on the initial Rayleigh number. The distance has a maximum around the values of $\text{Ra}$ at which the plumes transition to a turbulent-like state, displaying higher fluctuations. In this range of $\text{Ra}$ we also observed a change in morphology, with a cavity forming on the upper face of the object below.

The intricacy of the studied problem 
partially stems from the fact that the cold plume impinging on the bottom object originates from a source which is time dependent, and not uniform in space. Indeed, the ``mixed convection'' regime under which the bottom object melts is not just the result of a uniform incoming flow of a given velocity interacting with the natural convection generated by the melt water. 
In spite of this, we showed that a description considering the competition of a steady source for forced convection with natural convection is able to collapse the average Nusselt number across all of our simulations, even when ignoring the intricate time dependence of the plume originated from the top object, and the overall non-stationarity of a melting problem.
The competition between the shielding of the cold water and the enhanced mixing resulting from higher velocity fluctuations what underlays the observed dynamics, which, depending on the Rayleigh number, results in non-trivial morphological changes of the ice.

In view of this, it would be interesting to conduct a study imposing a uniform flow at a given speed and temperature, closer to a regime of mixed convection (i.e., not forced-convection dominated, as done by \cite{Yang2024_JFM}), to analyse whether these features are sufficient to observe a similar dynamics as the one observed for the lower object. This could provide insights on how to further disentangle the different contributions of temperature difference and velocity to the overall melting rate. 
Future studies could investigate how relevant the temporal dependence of the problem is in the observed dynamics, taking into account the effect of the frequency of oscillation of the melt water plume impacting on the lower object.

Furthermore, to compare with experiments, it would be of interest to perform a study in 3D. Even though the plume structure would be modified as compared to the two dimensional case, we expect to still recover, at least qualitatively, the enhanced and suppressed melting of the lower object. Additionally, evaluating the presence of salt in the ambient and varying the water temperature are interesting effects to explore that can contribute in going towards more natural, realistic scenarios.

\backsection[Supplementary data]{\label{SupMat}
Movies showing the evolution of the temperature field for different inter-object distances, for $\text{Ra} = 2.3\times 10^4,~5.0\times10^6$, and $2.3\times 10^7$, are provided as supplementary material. We also provide movies of the single object dynamics, for the same $\text{Ra}$, as comparison.
}

\backsection[Acknowledgements]{We thank Quentin Kriaa and Rui Yang for fruitful discussions.}

\backsection[Funding]{This work has been  funded by the European Union (ERC - Starting Grant, MeltDyn, No. 101040254, ERC - Advanced Grant, MultiMelt, No. 101094492). Numerical resources were provided by the EuroHPC Joint Undertaking for awarding the projects EHPC-REG-2022r03-208 and EHPC-REG-2023R03-178 to access the EuroHPC supercomputer Discoverer, hosted by Sofia Tech Park (Bulgaria). We also acknowledge the Dutch national e-infrastructure with the support of SURF Cooperative.}

\backsection[Declaration of interests]{ The authors report no conflict of interest.}

\backsection[Data availability statement]{The data that support the findings of this study are available upon reasonable request. 
}

\backsection[Author ORCIDs]{S. Angriman, https://orcid.org/0000-0003-2390-2800; D. Lohse, https://orcid.org/0000-0003-4138-2255; R. Verzicco, https://orcid.org/0000-0002-2690-9998; S.~G. Huisman, https://orcid.org/0000-0003-3790-9886}



\appendix

\section{Resolution convergence study}
\label{sec:grid_convergence}

We performed a resolution convergence study.
For an object with initial Rayleigh number $\text{Ra} \approx 2.90 \times 10^6$ ($L = 2.5~\text{cm}$), and considering an initial inter-object distance $D/L = 0.8$, we perform several simulations varying the number of grid points $N$. Here $N$ is the number of grid points used in the vertical direction, while in the horizontal direction $2N$ points are used. The phase-field variable is solved in a refined mesh, using twice as many divisions in both directions, i.e. $N_r = 2 N$.
To quantify only the effect of varying the grid resolution, and neglect possible variations due to changes in the initial condition, for the grid convergence study we turn off the noise in the initial temperature field.

\begin{figure}
\centering
\includegraphics[width=1\textwidth]{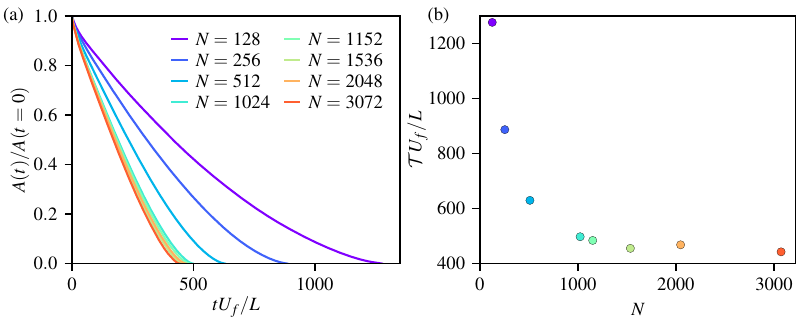}
\hfill
\includegraphics[width=1\textwidth]{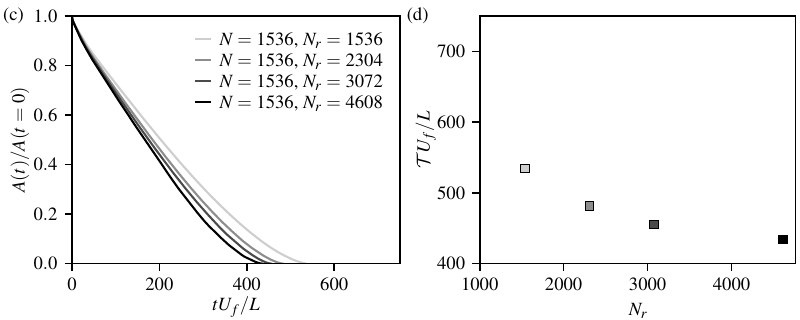}
  \caption{
  (a) Normalised evolution of the ice area as a function of time for different coarse grid resolutions. The resolution of the refined mesh $N_r$ is $N_r = 2N$.
  (b) Final melting time as a function of the grid resolution $N$. Colours are the same as in panel (a).
  (c) Evolution of ice area as for different refined grid resolutions, where $N=1536$ for all cases.
  (d) Final melting time as a function of the grid resolution $N_r$. Legend is the same as in panel (c). Note the differences in scales between panels (a), (c) and (b), (d).
  }
\label{fig:grid_convergence}
\end{figure}

Figure~\ref{fig:grid_convergence}(a) shows the evolution of the area of the top object as a function of time, for increasing number of grid points. The final melting time $\mathcal{T}$ for each case is shown in panel (b) of figure~\ref{fig:grid_convergence}.
The data show convergence for $N \geq 1152$. We use $N=1536$ for this Rayleigh number. Note we observe a similar trend for the lower object.
We also consider the effect of varying $N_r$ while keeping $N=1536$ fixed, shown in panels (c) and (d) of figure~\ref{fig:grid_convergence}, respectively. The data show convergence for $N_r \geq 2N$.

\section{Evolution of area and Nusselt number}
\label{sec:area_nusselt}

\begin{figure}
\centering
\includegraphics[width=1\textwidth]{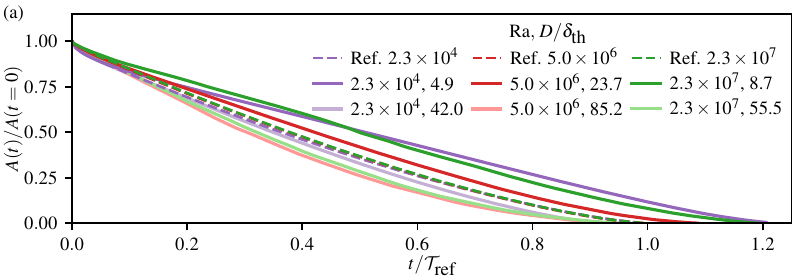}
\hfill
\includegraphics[width=1\textwidth]{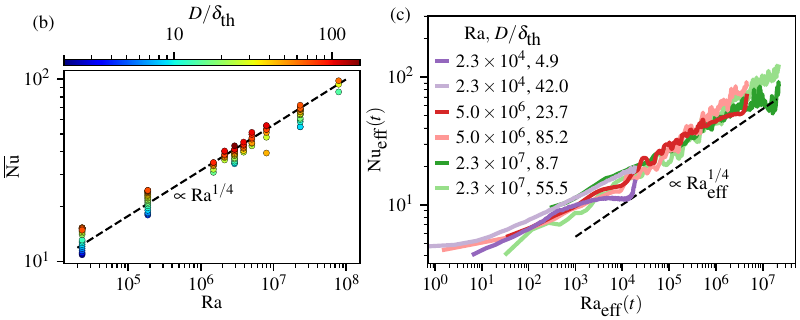}
\hfill
  \caption{
  (a) Normalised evolution of the area of the bottom object as it melts, for three initial values of $\text{Ra}$. The solid lines correspond to a small and a large initial inter-object distances. The dashed line shows the temporal evolution of the area of the single object case.
  (b) Average Nusselt number for the bottom object, estimated from the initial size and the melting time, as a function of the initial Rayleigh number for all of the datasets.
  (c) Effective Nusselt for the lower ice body and effective Rayleigh number, both computed from the instantaneous area, for the same datasets shown in panel (a).
  }
\label{fig:Nu_vs_Ra}
\end{figure}

The temporal evolution of the area $A(t)$ of the bottom ice object is shown in figure~\ref{fig:Nu_vs_Ra}(a), for three different initial sizes, $\text{Ra}$: $2.3\times 10^4$, $5.0\times 10^6$, and $2.3\times 10^7$ (corresponding to $L=0.5\,\text{cm},~3\,\text{cm}$, and $5$\,cm; we observe similar trends for the other datasets). For each object size, we show $A(t)$ for the reference case, and for a small and a large inter-object distance $D$ (where melting is delayed and enhanced, respectively). Considering that the top object behaves like the reference case, from the figure it can be seen that the top/bottom asymmetry appears early in the evolution of the system. That is to say, for the large inter-object distances, where the final melting times are such that $\mathcal{T}_\text{bot} < \mathcal{T}_\text{top}$, it also occurs that at a given intermediate time $t_0$, $A_\text{bot}(t_0) < A_\text{top}(t_0)$. We observe the corresponding inverse behaviour for small inter-object distances where $\mathcal{T}_\text{bot} > \mathcal{T}_\text{top}$.

The Nusselt number is defined as the ratio of total heat transfer through the ice interface to conductive heat transfer
\begin{equation}
    \text{Nu} = \frac{L}{\Delta T} \frac{\partial T}{\partial n}\Big\lvert_\text{interface}.
    \label{eq:Nu}
\end{equation}
On the other hand, for a solid at constant temperature the Stefan boundary condition determines that the normal recession velocity of the solid interface $\mathbf{u}\cdot\hat{\mathbf{n}}$ is proportional to the temperature gradient at the interface \citep{Worster2000},
\begin{equation}
    \mathbf{u}\cdot\hat{\mathbf{n}} = \frac{\kappa\, c_p}{\mathcal{L}} \frac{\partial T}{\partial n}\Big\lvert_\text{interface}.
    \label{eq:Stefan_BC}
\end{equation}
An average recession velocity of the interface can be estimated as $u = L/\mathcal{T}$, and then combining this with equations $\eqref{eq:Nu}$ and $\eqref{eq:Stefan_BC}$, the melting time $\mathcal{T}$ and the time evolution of the area can be related to the surface-averaged heat flow through an average Nusselt number $\overline{\text{Nu}}$
\begin{equation}
    \overline{\text{Nu}} = \frac{L^2/\kappa}{\mathcal{T}}\,\frac{1}{\text{Ste}}.
\end{equation}
Re-writing this expression in terms of the input parameters of the system reads
\begin{equation}
    \frac{L/U_f}{\mathcal{T}} = \frac{\text{Ste}}{\text{Pr}^{1/2}\, \text{Ra}^{1/2}}\, \overline{\text{Nu}}.
\end{equation}

From this expression we can estimate an average Nusselt number from the melting time, and compare it with the Rayleigh number $\text{Ra}$. Figure\,\ref{fig:Nu_vs_Ra}(b) shows $\overline{\text{Nu}}$ as a function of $\text{Ra}$, computed from the melting time of the bottom object, for all of the initial inter-object distances explored. We show also a power law with exponent $1/4$ for reference, and observe that the data follow the scaling reasonably  well, consistent with what is observed for heat transfer in natural convection \citep{Bejan1993}.

\begin{figure}
\centering
\includegraphics[width=1\textwidth]{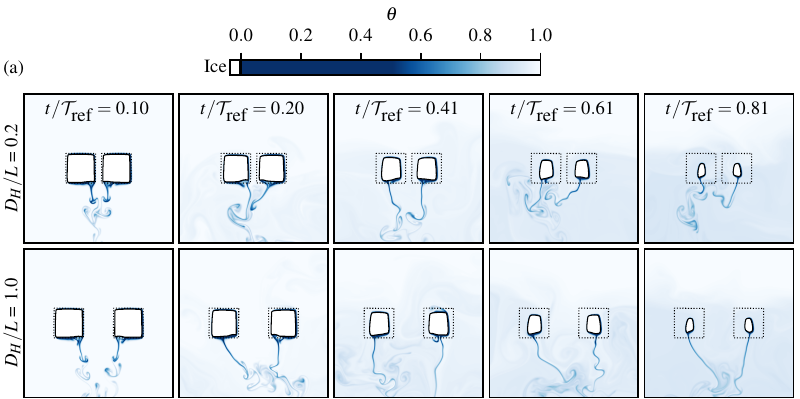}
\hfill
\includegraphics[width=1\textwidth]{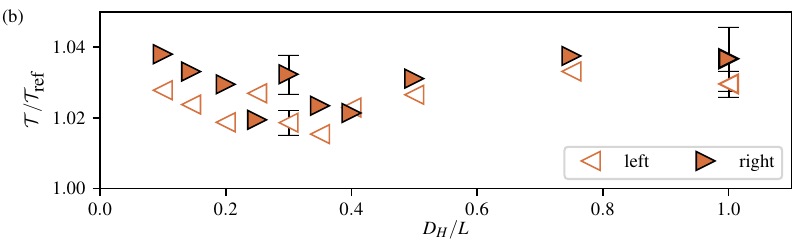}
\hfill
  \caption{Ice objects separated horizontally a distance $D_H$, with initial size $L
  = 5~\text{cm}$ ($\text{Ra} \approx 2.3\times 10^7$).
  (a) Zoomed-in evolution (see figure~\ref{fig:setup}a) of the melting of the left and right objects. The dotted lines indicate the initial contours. 
  The instantaneous snapshots of the non-dimensional temperature field $\theta$ are shown for different times, in units of the reference case melting time $\mathcal{T}_\text{ref}$. 
  The top row corresponds to an initial separation $D_H/L = 0.2$, while the bottom row is for $D_H/L = 1.0$.
  (b) Melting times $\mathcal{T}$ of the left and right objects in units of the single object case melting time $\mathcal{T}_\text{ref}$, as a function of the normalised horizontal distance between objects.
  }
\label{fig:horizontal}
\end{figure}

To check if this scaling holds during the evolution of the system, we can define an effective Nusselt number $\text{Nu}_\text{eff}(t)$ from the time derivative of the object's area,
\begin{equation}
    \frac{1}{U_f L} \frac{dA(t)}{dt} = \frac{\text{Ste}}{\text{Pr}^{1/2}\, \text{Ra}^{1/2}}\, \text{Nu}_\text{eff}(t),
\end{equation}
which we compare with an effective Rayleigh number based on the instantaneous object size, i.e. $\text{Ra}_\text{eff}(t) = \text{Ra}\, \frac{A^{3/2}(t)}{L^3}$. This is shown in figure~\ref{fig:Nu_vs_Ra}(c), for the bottom object and for the same datasets as those shown in panel (a) of figure~\ref{fig:Nu_vs_Ra}. We observe again a power-law-like behaviour, with an exponent close to $1/4$, which is shown as a reference.


\section{Melting of horizontally separated ice objects}\label{sec:horizontal}

We also studied the effect of placing two ice bodies side by side horizontally, separated a distance $D_H$. We follow the same techniques as previously described, and consider the same conditions as those shown in figure~\ref{fig:setup}(a), except the objects are located at $y=5\,L$.
We explore the melting dynamics of objects with 
a Rayleigh number $\text{Ra} \approx 2.3\times 10^7$ ($L = 5~\text{cm}$).
Figure~\ref{fig:horizontal}(a) shows instantaneous snapshots of the normalised temperature field as the bodies melt, for two distances $D_H/L = 0.2$ and $D_H/L=1$, in the top and bottom rows, respectively. Time is again compared with the melting time $\mathcal{T}_\text{ref}$ of a single object in the same domain. We observe that the melting occurs slightly faster on the inner side of the objects for both $D_H$ shown, likely due to enhanced circulation resulting from the melt water plumes interacting with each other. However, there does not seem to be a significant difference in the morphology of the objects for the two separations shown.
When computing the melting time $\mathcal{T}$ for $D_H/L \in (0.1,1]$, shown in figure~\ref{fig:horizontal}(b) we observe that the melting times are of the order of $3\%$ larger than the single object case, differences that do not seem to be statistically significant. Indeed, the error bars shown for some data points in panel b of figure~\ref{fig:horizontal} indicate one standard deviation of $\mathcal{T}$ over $10$ different realisations of the same configuration, by varying the noise in the initial temperature field (following the same procedure as for the vertical separation study).
We do observe a weak trend in $\mathcal{T}$ with $D_H$, with differences of about $2\%$, possibly related to the melt water plumes interacting slightly more with each other for smaller $D_H$. No meaningful differences in the mean circulation can be observed between the different cases.

\section{Additional correlations}
\label{sec:correlations}

\begin{figure}
    \centering
    \includegraphics[width=1\linewidth]{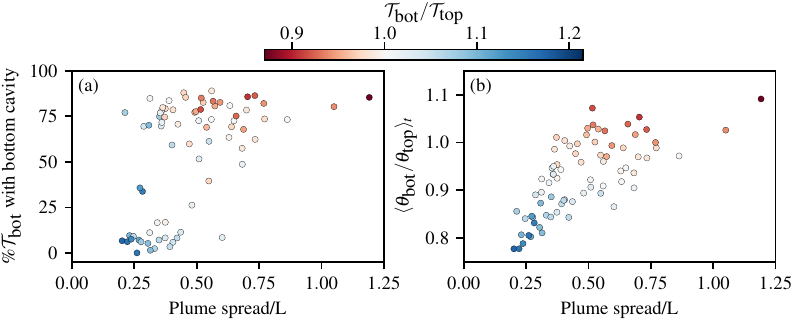}
    \caption{
    (a) Percentage of the bottom object evolution where a cavity is detected, and (b) ratio of bottom-to-top temperatures, as a function of the plume spread in units of $L$, for cases with $\text{Ra}> 10^6$. The colour of the markers indicates the ratio of melting times $\mathcal{T}_\text{bot}/\mathcal{T}_\text{top}$, as indicated by the colour bar.
    }
    \label{fig:correlations}
\end{figure}

We quantify the link between the spread of the plume from the reference object, and the observed morphology and temperature in the surroundings of the bottom object, for the cases with $\text{Ra} > 10^6$. 
Figure~\ref{fig:correlations}(a) shows that a persistent bottom cavity is linked with a plume spread of at least $L/2$. These points also correspond to the enhanced bottom object melting (red data).
Panel (b) of figure~\ref{fig:correlations} shows that the more the plume is spread horizontally, the warmer the surroundings of the bottom object are as compared to the top object, namely a wider plume entrains more warm ambient water.


\bibliographystyle{jfm}
\bibliography{ms}

\end{document}